\documentclass[fleqn,10pt]{wlscirep}
\usepackage[utf8]{inputenc}
\usepackage[T1]{fontenc}
\title{Evolution beats random chance: Performance-dependent network evolution for enhanced computational capacity}

\author[1,*]{Manish Yadav}
\author[2]{Sudeshna Sinha}
\author[1]{Merten Stender}
\affil[1]{Chair of Cyber-Physical Systems in Mechanical Engineering, Technische Universität Berlin, Straße des 17. Juni, Berlin, 10623, Germany}
\affil[2]{Department of Physical Sciences, Indian Institute of Science Education and Research Mohali, Sector 81, SAS Nagar, 140306, Punjab, India}

\affil[*]{manish.yadav@tu-berlin.de}

\begin{abstract}
The quest to understand \textit{structure-function} relationships in networks across scientific disciplines has intensified. However, the optimal network architecture remains elusive, particularly for complex information processing. Therefore, we investigate how optimal and specific network structures form to efficiently solve distinct tasks using a novel framework of performance-dependent network evolution, leveraging reservoir computing principles. Our study demonstrates that task-specific minimal network structures obtained through this framework consistently outperform networks generated by alternative growth strategies and Erdős-Rényi random networks. Evolved networks exhibit unexpected sparsity and adhere to scaling laws in node-density space while showcasing a distinctive asymmetry in input and information readout nodes distribution. Consequently, we propose a heuristic for quantifying task complexity from performance-dependently evolved networks, offering valuable insights into the evolutionary dynamics of network \textit{structure-function} relationship. Our findings not only advance the fundamental understanding of process-specific network evolution but also shed light on the design and optimization of complex information processing mechanisms, notably in machine learning.
\end{abstract}

\begin{document}

\flushbottom
\maketitle
%
%
\thispagestyle{empty}

\noindent\textbf{Keywords:}{ evolving networks, reservoir computing, performance-dependent evolution, enhanced information processing, network structure-function}


\section*{Introduction}

Naturally occurring networks process diverse and complex information with limited resources. Biological networks such as neuronal, protein interaction, and gene regulatory networks (GRNs) are the best examples of large-scale complex systems that have grown and evolved by natural selection over millions of years to \textit{efficiently} process complex biochemical signals \cite{BrainStrFun_BassettReview, GRN_Evolution_Babu, PIN_Evolution}. These networks exhibit characteristic structural organizations with distinct features such as scaling laws, resulting in robust and efficient functionality-specific information processing \cite{EfficientBrain_BassettPLOS, BrainStrFun_BassettReview}.

The growth and evolution of networks have been a central topic of research in the network science community that has provided several scaling laws and unique graph theoretic properties over the years \cite{EvolNet_Review_2002}. For instance, the well-known Price's preferential attachment model introduced in the 1960s \cite{PricesModel} provided an understanding of the power-law degree distribution of growing co-citation networks over time. However, the main research focus has remained around the parameter-dependent self-organization of networks showcasing their statistical growth and edge reorganization \cite{SmallWorld, NetEvolNature_2023}. The focus of many disciplines in recent years has shifted to the \textit{structure-function} relationship of large complex networks. For instance, the relationship across distinct regions and connectivity was found to scale in the brain  \cite{BrainStrFun_BassettReview, BrainStrToFun_Yanchuk2022, BrainStrToFun_SUAREZ2020, BrainStrToFun_RNN_2023, BrainStrToFun_Ex_Review2016}, and recurrent structures facilitate the memory-dependent temporal information processing in GRNs of \textit{Escherichia coli} \cite{EColi_Recerrence_2004}, and in several other evolutionary distant species \cite{RC_Jordi}. The adaptive dynamic networks theory is a great step toward understanding the functional relationship of a network to its structural organization \cite{AdaptiveNetReview_Yanchuk_2023}. This theory covers networks that can change their connectivity over time according to their dynamical state. However, the question of how these large-scale natural networks evolve to achieve specific functionality and what should be their natural structural organization to exhibit a function is still missing a general conceptual understanding and explanation.

A different perspective on this problem comes from artificial neural networks (ANNs), which are inspired by biological networks and developed to have specific functionalities for wide-ranging tasks, such as forecasting dynamics. However, the lack of a formal theory explaining process-specific network \textit{structure-function} relationship leaves the researchers with randomly selecting the size and structures of the artificial networks, a process that is mostly based on a trial-and-error strategy or one's prior experiences \cite{ANN_HitNtrial_Review}. There is no guiding principle for network structure selection that can efficiently solve a given task with given resources. Increasing the network size is often attempted in hopes of eventual task-solving\cite{ANN_LargeSize_Review}. It would be of utmost value if there was some framework in which to gauge ``how big is big enough'' in a general scenario. More importantly, from the fundamental point of view, it is of immense interest to uncover underlying concepts that can lead to the smallest effective networks for information processing. 

Motivated by the importance of the problem across multiple disciplines as well as its valuable applications, we are describing the formation of \textit{function}-specific network \textit{structures} in this paper. For that, we propose a general framework of process-specific network evolution that will help elucidate not just the network \textit{structure-function} relationship from an evolutionary perspective but also in generating task-specific artificial networks. Inspired by biological systems that can solve various tasks with limited resources, we hypothesize that one can reach functionality-specific optimally sized networks more effectively through evolutionary concepts, rather than by random chance. For that, we propose the idea of a performance-informed growth strategy that lets networks naturally evolve into specific and efficient structures for solving a given task with desired performance. This framework will eventually help in the identification of broad principles to grow networks, starting from a few seed nodes, which can potentially allow the synthesis of small networks with enhanced computational capacity.

The proposed idea of functionality-specific network formation can be implemented with an information-processing framework that operates on any network irrespective of its architecture, number of nodes or edges. Such a framework will facilitate the underlying information-processing network with the freedom to grow, morph and evolve into functionality-specific size and architecture. The Reservoir Computing (RC) framework matches such criteria where input signals are mapped into a higher dimensional structure-free \textit{reservoir} \cite{RC_Review2019, RC_book2021}. RC is a generalization of earlier neural network architectures such as echo-state networks and liquid-state machines, derived from recurrent neural networks initially introduced to explain information processing in neuronal networks \cite{Maass2002, Jaeger2001}, and this framework has sparked intense wide-ranging interest in recent years \cite{RC_net,RC_soliton,RC_edgechaos,RC_circuit,pendulum_RC,RC_spatiotemporal,RC_Hamiltonian,RC_optical}. An RC has 3 layers namely the input, the \textit{reservoir} and the output layer. A dynamic input signal $\textbf{u}(t)$ is fed into the reservoir, and then a simple readout mechanism is trained to obtain a mapping to the desired output. The dynamics of the \textit{reservoir} is iteratively obtained for time $t+1$ using the Eq. \ref{eq:RC1}:

\begin{align}
\textbf{r}(t+1) &= (1-\alpha)\textbf{r}(t) + \alpha L\Bigl(W^{res}\textbf{r}(t) + W^{in}\textbf{u}(t)\Bigl) \label{eq:RC1}\\
\Tilde{\textbf{y}}(t) &= W^{out}\textbf{r}(t) \label{eq:RC_train}
\end{align}

where $W^{res}\in\mathbb{R}^{N\times N}$ is the adjacency matrix representing all the connections between ($N$) reservoir nodes. $W^{in}$ specifies which nodes of the reservoir network receive the time-varying input. Each node's dynamics are defined by the previous state, external stimuli and nonlinear activation function $L$. The reservoir matrix $\textbf{r}(t)\in\mathbb{R}^{N}$ contains information of the reservoir states at a given time $t$ when it is stimulated by an external signal $\textbf{u}(t)\in\mathbb{R}^{T}$. Parameter $\alpha\in[0,1]$ reflects the reservoir information leakage and determines the amount of the past reservoir information being leaked over time. The RC approximated real-time output $\Tilde{\textbf{y}}(t)$ is obtained using a linear combination of the reservoir states given by the Eq. \ref{eq:RC_train}. A key feature of a reservoir computer is that only the readout is trained while the reservoir network stays fixed. The output weight matrix $W^{out}$ is trained to minimize the difference between the RC output and the actual output $\textbf{y}(t)$ which is typically obtained using the Ridge regression, also commonly known as \textit{one-step training}.

Reservoir networks are typically chosen to have a random structure with hundreds to a few thousands of nodes \cite{RC_Review2019, RC_book2021}. Since there are no specific rules for selecting the network topology of a reservoir, the RC framework provides an opportunity to study the computational capabilities of self-organizing and growing networks vis-a-vis a Erd\"os-Renyi random networks employed as a reservoir. Therefore, the general problem of network \textit{structure-function} relationship can be studied via the RC framework which will be beneficial for the research community exploring the information processing behavior of RCs as complex dynamical systems. Moreover, a growing number of research has computationally as well as experimentally proved that the \textit{liquid state machines} or RC paradigm can explain the processing of complex biological computations through recurrences 
 \cite{Maass2002, RC_Jordi, BioComputWithRecurrences_Jordi2024, RC_withConnectome_PLOS2022, RC_BioComputing_Science2024, RC_withConnectome_NatComm2024}. This suggests and provides another argument to directly employ a recurrent network-based framework to study how networks grow and evolve to achieve function-specific structures \cite{RC_forEvolution_RoyalSociety2019}. Therefore, all the questions that we posed above for understanding the network \textit{structure} dependent \textit{functionality} can be directly translated to that of \textit{reservoir} network. Our central idea is to implement an evolving network model into the RC paradigm and observe the graph-theoretic properties of the reservoir networks as they evolve while solving different tasks. The principal focus will be the following: 
 Is it possible to generate optimally sized networks to solve specific tasks? If yes, what is the comparative size of an evolved network that delivers the same performance as a random reservoir? What are the graph theoretic structural differences between evolved networks and random reservoirs? Are there any unifying underlying features in evolved networks grown for different tasks?

In this work, we propose a novel framework showcasing the functionality-specific network generation via their performance-dependent evolution (see schematic in Fig.\ref{fig1}). We demonstrate that the evolved networks are significantly smaller than their randomly connected counterparts, suggesting that evolutionary strategies converge to more parsimonious computing structures than random chance. We also suggest a different set of evolution strategies and explore their comparative performances. Further, we uncover scaling relations and graph-theoretic features of the evolved networks marking the formation of optimally sized and computationally efficient network structures across different tasks. These special features have the potential to be used as prescriptive recipes for designing optimal machine-learning algorithms consistent with the scaling relations.

\section*{Results}

\subsection{Performance-dependent network evolution framework}

The key idea of generating minimal networks for efficient processing is based on the evolutionary growth of the networks. Such performance-enhancing growth can be achieved by adding nodes to a pre-existing network, as long as they improve the overall information-processing capabilities of the network. Nevertheless, a node improving the performance of the system can also become obsolete in later evolution stages when it no longer serves any useful functionality within the network, e.g. when encoding redundant information processed by other nodes. Therefore, node deletion plays a crucial role, as the presence of redundant nodes makes the entire system inefficient in terms of, for instance, energy, space, and computation time. Therefore, to achieve an evolution-like growth of the networks, the network evolution framework should include a mechanism to continuously {\em add and delete nodes} to/from a pre-existing network. That framework needs to be embedded into a feedback loop with the specific task to solve, to inform the evolution of the information processing capabilities of the network.

\begin{figure}[h]%
\centering
\includegraphics[width=0.8\textwidth]{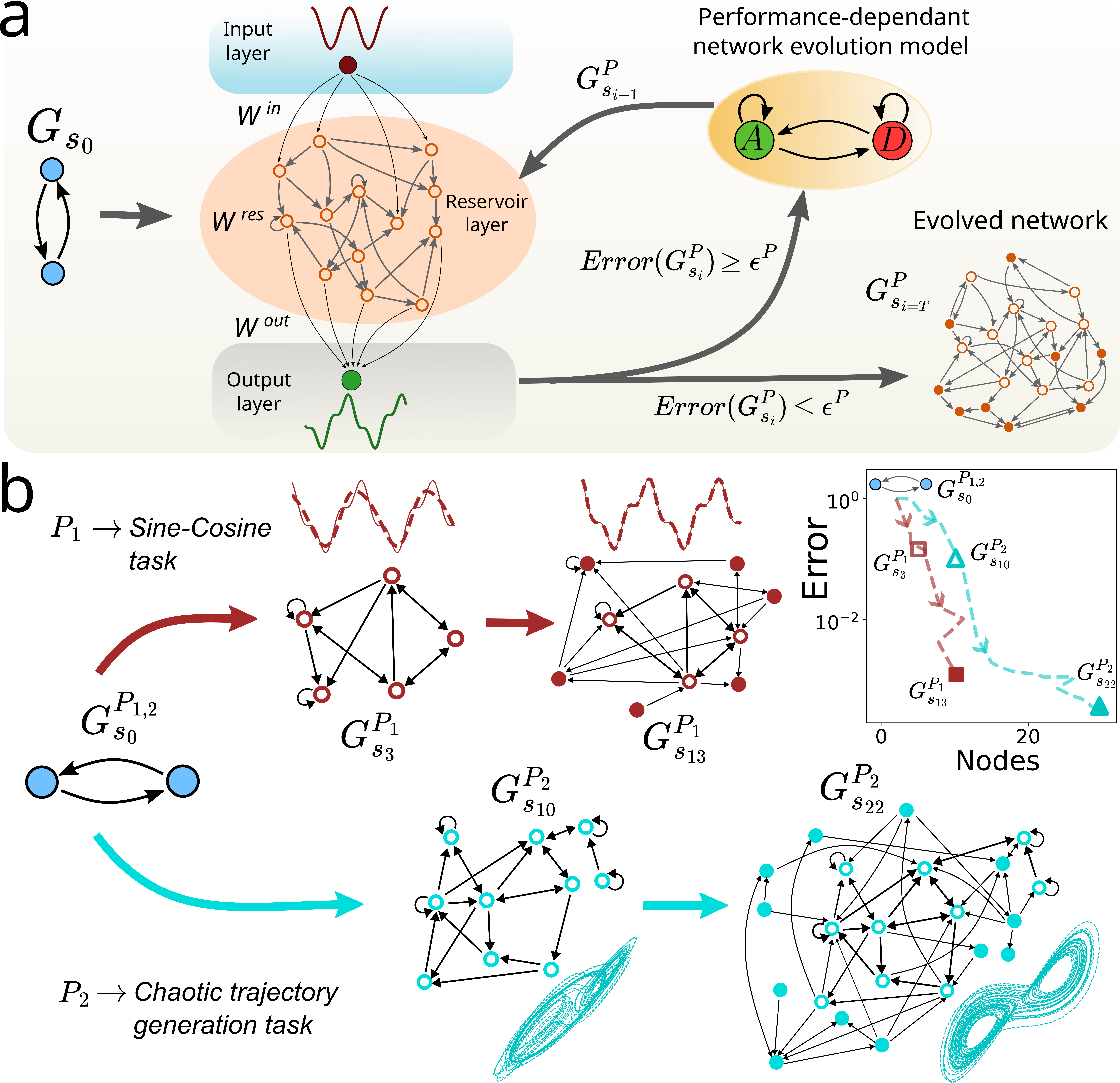}
\caption{
\textbf{Performance-dependent network evolution.} (a) Schematic showing the working mechanism of the network evolution model depending on an underlying process or task ($P$). The model consists of a node addition module ($A$) and a node deletion module ($D$). Module $A$ adds a new node to the pre-existing network while module $B$ removes nodes from the pre-existing network, conditioned on network performance improvement. This procedure continues until the desired performance ($\epsilon^{P}$) is achieved. Starting with a very small initial network ($G_{s_{0}}$), the model generates an evolved and minimal final network ($G^{P}_{T}$) that can efficiently solve the given task $P$. (b) Schematic showing the model evolving the same initial network $G_{s_{0}}$ for two different exemplar tasks; $P_{1}$ for converting the sine function to its complex conjugate and $P_{2}$ for generating a chaotic trajectory from the Lorenz system. The final evolved networks ($G^{P_{1}}_{s_{18}}$ and $G^{P_{2}}_{s_{22}}$) contain some nodes (hollow) from a network at some previous intermediate stage ($G^{P_{1}}_{s_{3}}$ and $G^{P_{2}}_{s_{10}}$). The evolved networks can solve the underlying task with higher accuracy as compared to their intermediate networks.
}\label{fig1}
\end{figure}

Pursuing this line of thought, we propose a performance-dependent network evolution framework that is provided with \textit{node addition ($A$)} and \textit{deletion ($D$) modules} interconnected with each other in a feedback loop, schematically shown in Fig. \ref{fig1}(a). The primary function of module \textit{A} is to add new nodes to the pre-existing network in such a way that will improve the information processing capabilities of the new network. On the contrary, the module \textit{D} continuously removes the unnecessary nodes from the network. A minimal two-node network $G_{s_{0}}$ is considered as an initial network at the beginning of the evolutionary iteration $s_{0}$ throughout this study. This minimal network reaches the evolution modules $A$ and $D$ initiating the task-specific evolutionary growth. A final evolved network $G^{P}_{s_{T}}$ is generated when a desired performance ($\epsilon^{P}$) for an underlying task $P$ is achieved at iteration $s_{i=T}$. In a nutshell, the \textit{modules} $A$ and $D$ will orchestrate the growth of the network in such a way that will improve its information processing capabilities by working together to retain only those nodes that assist in enhancing its overall performance while eliminating the ones that do not (Fig. \ref{fig1}(b)).

The network evolved by the performance-dependent network evolution model is considered as the reservoir layer network ($W^{res}$) of the RC framework at every evolutionary iterative step ($s$). The reservoir networks will evolve to achieve different functionalities embarked with a range of distinct underlying processes of varying levels of complexity ranging from classical one-to-one mappings, and multidimensional trajectory generation to memory-dependent tasks. In total, we considered seven different tasks: two Sine-Cosine tasks, NARMA-5th, 10th and 15th order tasks \cite{NARMA1}, chaotic trajectory prediction of a prototypical chaotic system and phase-portrait generation of the limit cycle from a nonlinear relaxation oscillator.

\begin{figure}[h]%
\centering
\includegraphics[width=1\textwidth]{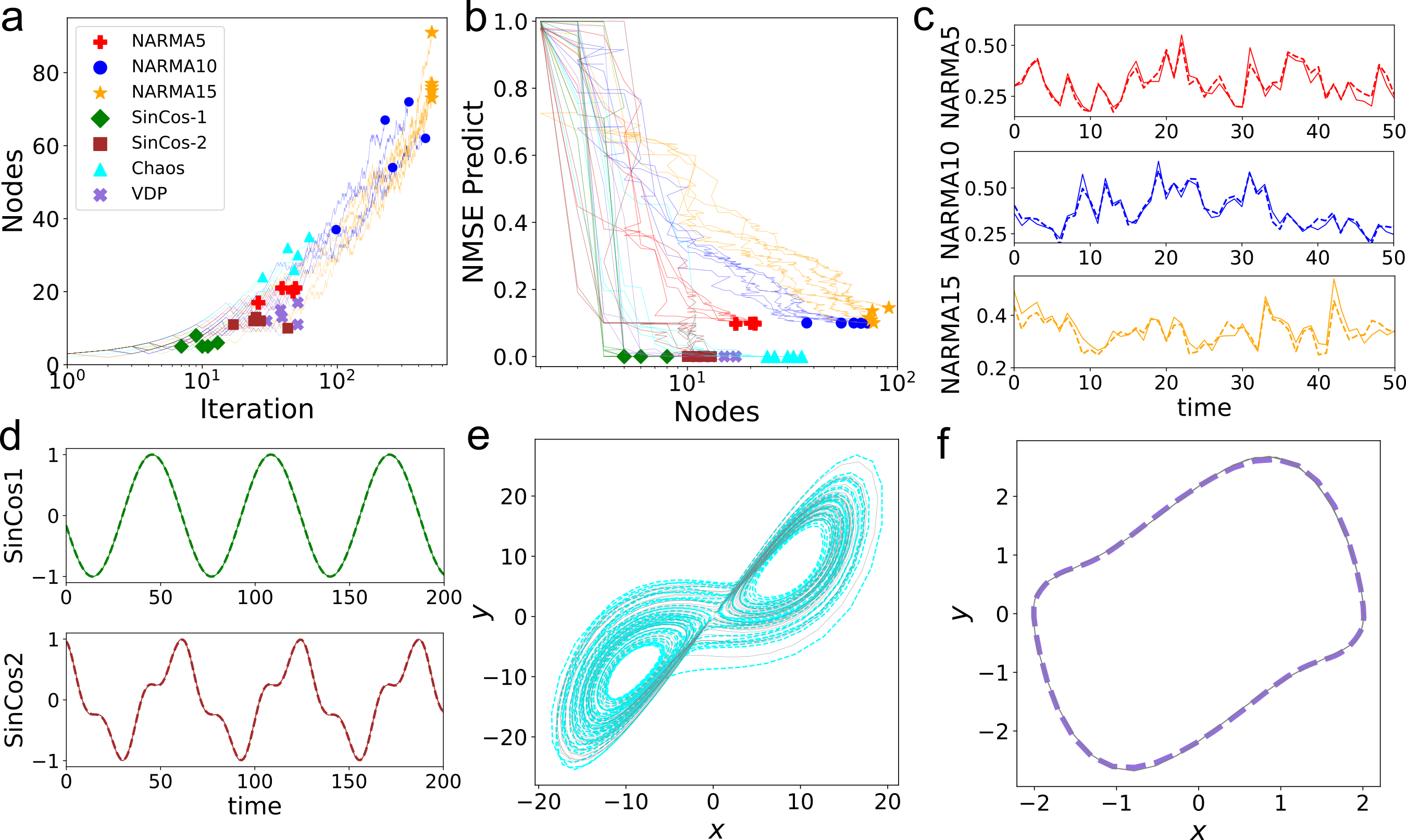}
\caption{
\textbf{Generalization of the performance-dependent network evolution over diverse benchmark tasks.} (a) The number of nodes in the evolving networks is shown over each iteration of the evolution model for different tasks. The total number of nodes in the final evolved networks is represented with different symbols. (b) The performance (Normalized Mean Square Error, NMSE) for the independently evolving networks is represented against the total number of nodes present in the corresponding evolving network. The evolution of 5 representative networks for each task are displayed in (a) and (b). (c) Predicted outputs generated by the evolved networks for the three different NARMA time series prediction tasks of 5th (top), 10th (middle) and 15th (bottom) orders, are plotted with dashed curves and the true values are with solid ones. (d) Predictions of SinCos-1 and SinCos-2 tasks. (e) Chaotic trajectories from the Lorenz system and (f) phase-portrait of the limit-cycle from the Van der Pol oscillator obtained from evolved networks for the respective cases. RC is working autonomously in a closed-loop manner in (e) and (f). 
}\label{fig2}
\end{figure}

The proposed performance-dependent evolution model has been used to solve the aforementioned tasks. The evolution of the size of the networks for different tasks is represented by the number of nodes over evolutionary iterations in Fig.~\ref{fig2}(a). Furthermore, the performances as reflected by the Normalized Mean Square Error, denoted by NMSE, of these evolving networks are represented against the number of nodes in Fig.~\ref{fig2}(b). The model takes a distinct number of iterations depending on each task (Fig.\ref{fig2}(a)) and stops when the underlying performance ($\epsilon^{P}$) is achieved. The final evolved networks (represented with different markers) for distinct tasks are different from each other in terms of size, marked by the total number of nodes. It is evident that the Sine-Cosine-1 is the least complex task of mapping the Sine to a Cosine function that can be solved with the least complex networks, some consisting of just 5 nodes. A slightly more complex task of mapping the Sine function to a complex polynomial of Sine and Cosine functions, namely the Sine-Cosine-2 task can also be solved with small networks of just 11 nodes on average. 

The networks evolved to solve the input memory-dependent tasks of NARMA-$5$, $10$ and $15th$ order have average sizes of 20, 55 and 75 respectively. This suggests that the NARMA-type tasks are more complex than Sin-Cosine tasks as they require bigger recurrent networks to achieve specific memory requirements to solve them. Furthermore, it is worth noting that the evolved reservoir networks for NARMA-type tasks are orders of magnitude smaller than typically used, showing that memory-dependent tasks can also be solved by smaller networks depending on the memory requirements. Sample time-series predictions from NARMA and Sine-Cosine tasks from one of the evolved networks are respectively represented in Fig.\ref{fig2} (c) and (d). NARMA-15 requires the most number of nodes out of all the benchmarks and NARMA tasks which reflects its high complexity. In almost all the cases for the NARMA-$15$ task, the total network evolutionary iterations limit of the model $S^{Evolve}$ was reached and the final performance was slightly lower than the requisite  $\epsilon^{{NARMA-15}}$ i.e $E(G_{S^{Evolve}}^{NARMA-15})\geq\epsilon^{{NARMA-15}}$. The tasks for multidimensional phase-space trajectory generation have been made more complicated as they have been generated by autonomously working reservoir computers for both tasks. 
Networks evolved for generating the limit-cycle of the Van der Pol oscillator contained an average of just 12 nodes while those of the 3D chaotic trajectory of the Lorenz system contained only 27 nodes, resulting trajectories are shown in Fig.\ref{fig2} (e) and (f) respectively.

From this initial study, it is therefore evident that the performance-dependent evolution model demonstrates the existence of small networks with just a few nodes that can efficiently process complex tasks. Another interesting observation is that after starting from the same initial two nodes, the networks evolve to different sizes that are optimally structured to solve specific tasks with different complexity levels, also schematically depicted in Fig.\ref{fig1} (b). Simultaneously, networks evolved to carry out the same process are clustered together with a slight deviation in their sizes. These results are consistent with the observations in naturally evolving networks, such as the presence of distinctly sized and structured brain regions for different functionalities \cite{BrainStrToFun_SUAREZ2020, BrainStrFun_BassettReview}. An example of the latter is the signaling networks responsible for regulating the same cellular processes which are found to be actively present in many different cell types and are conserved over evolutionary scale \cite{Conserved_SigNet}. Furthermore, these results are contrary to the common practice of employing randomly connected large networks, typically of size $N\sim\mathcal{O}(10^{2-3})$ to solve complex tasks using reservoir computers \cite{Jaeger2001, RC_2023_MemoryAug, RC_2020_SciRep}. Several questions arise from these initial observations: Is it necessary for the network to be grown and contracted continuously to achieve an efficient computational capacity and structure? Is performance-dependency necessary to generate computationally efficient network structures? How are these evolved networks different from the commonly employed randomly connected networks? Is there a scaling law that networks follow while evolving to solve a task?

\subsection{Challenging the performance-dependent network evolution with different growth strategies}

To answer the aforementioned questions and also, to challenge the results of our performance-based network evolution model, we considered two different network growth strategies. First, to elucidate the requirement of performance dependency for efficient network generation, an uninformed network growth model (N1) is considered as a reference. There will be no feedback to the model for the network performance meaning that the network growth in this model will depend completely on random chance irrespective of any task and performance. To understand the necessity of network growth and contraction a second alternate model is formulated only with growth or the performance-dependent network growth model (N2). The N2 model will only be able to expand a network depending on its performance but will not be able to remove nodes. It can be thought of as a partial network evolution model as it consists only of node \textit{addition module A}. From here onwards N3 will be used to refer to the principal performance-based network evolution model proposed in the Section above.

\begin{figure}[h]%
\centering
\includegraphics[width=0.99\textwidth]{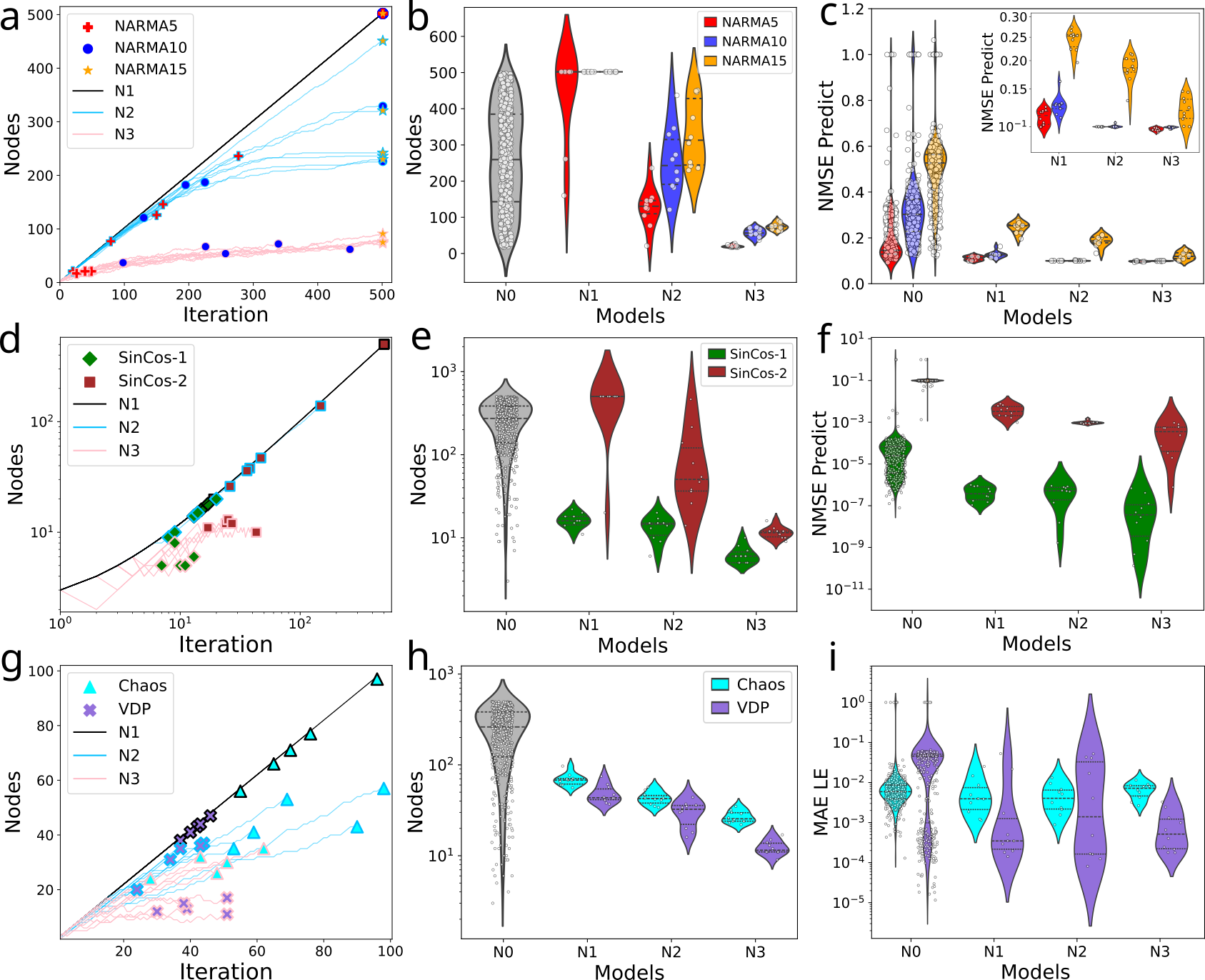}
\caption{\textbf{Contrasting networks generated by the performance-based network evolution with alternate network growth strategies.} Comparison of the total number of nodes as the networks grow/evolve in the models N1 (black curve), N2 (sky blue curves) and N3 (pink curves) for NARMA-5th, 10th and 15th order tasks in (a), for SinCos-1 and 2 tasks in (d) and Lorenz's chaotic trajectory with Van der Pol's limit cycle generation tasks in (g). The total number of nodes in the final evolved networks is represented with different symbols corresponding to each task. (b), (e) and (h) The nodes in the final evolved networks from N3 model are compared with that of N1 and N2 models along with Erd\"os-Renyi random networks (N0, grey). (c), (f) and (i) The performance calculated using Normalized Mean Square Error (NMSE) of 10 independently evolved/grown networks by N1, N2 and N3 models for all the tasks are compared alongside the performance of Erd\"os-Renyi random networks (N0). The Mean Absolute Error of the Lyapunov-Exponents (MSE-LE) between the original and predicted trajectories is used to quantify the error in both the trajectory generation tasks
 (i).}\label{fig3}
\end{figure}

Starting with the same 2-node initial network, N1 and N2 models were used to grow the networks for all the seven benchmark tasks. The network growth trajectories of the three models N1 (black), N2 (blue) and N3 (pink) for the NARMA-$5$, $10$ and $15th$ order tasks are displayed together for comparison in Fig.\ref{fig3} (a). The networks generated by the performance-independent N1 model grow linearly over the iterations until the desired performance ($\epsilon^{P}$) is achieved. In
almost all cases, N1 models reach the iteration limit $S^{Evolve} = 500$ for all the three NARMA
tasks that were set the same for all the growth models. Also in most cases, the final network did not manage to successfully execute the NARMA tasks (especially the more complex NARMA-15) to the prescribed accuracy. However, the N3 model not only achieved the prescribed accuracy but also generated optimal networks without reaching the maximum iteration limit $S^{Evolve}$. Similarly, for the Sine-Cosine (Fig.\ref{fig3} (d)), Lorenz's chaotic trajectory and Van der Pol's limit-cycle generation (Fig.\ref{fig3} (g)) tasks, the networks follow a linear growth for the N1 model, while a sublinear growth for the N2 and N3 models. All the networks evolved by the N3 model achieve the desired performance with the least number of iterations consisting of a lesser number of nodes which is consistently observed for all the benchmark tasks. A side-by-side comparison of the distinct network growth strategies (N1 and N2) demonstrates several key aspects of the performance-dependent network evolution (N3). Specifically, examining the final networks generated by the different models for different tasks reveals that the performance-dependent growth strategy generates not only the smaller networks with fewer nodes (Fig.~\ref{fig3} b,e,h) but these evolved networks also exhibit enhanced information processing capabilities reflected by the least error (Fig.~\ref{fig3} c,f,i).

There is a minimal bound to the number of nodes required for a certain task, irrespective of the network structures imposed by different evolution strategies as the initial 2-node network did not manage to solve any of the given tasks. This minimal bound is a proxy for the task complexity, i.e. a universal or task-agnostic quantity to characterize learning tasks which today has not been proposed elsewhere. Furthermore, if a network grows uninformed without any performance-dependent strategy, the number of nodes is expected to increase linearly with iterates. Another interesting observation is that while uninformed node addition (N1), does indeed on average, give rise to linear growth of nodes with iterations, performance-dependent node addition (N2) yields sub-linear growth of nodes. Additionally, the number of nodes needed to achieve a performance level saturates after a certain number of iterates, i.e. one obtains a converged network that does not need to be bigger to solve the task. This implies that not all additions are beneficial to performance, and an emergent network smaller than that obtained by the uninformed growth (N1) is sufficient. 

Further, the performance-dependent evolved networks allowing node addition as well as removal (N3) yield the slowest growth of nodes with iterations. Importantly, the number of nodes needed for the achievement of a prescribed level of performance is significantly lower for networks evolved using a performance-dependent addition and removal strategy (N3), often being an order of magnitude smaller than uninformed growth (N1). In this case, the number of nodes saturates very quickly to a small value. This leads to the important conclusion: these evolved networks can perform tasks efficiently and parsimoniously, with small networks yielding very low errors on a wide range of tasks.
Another distinguishing feature is that the spread in the number of nodes of the evolved network is the smallest for the performance-dependent network with node addition and removal. In contrast, if we consider a growing network only with node addition (N2), the variance in the number of nodes of the converged network is very high. This direct comparison of distinct network growth strategies therefore answers that it is indeed necessary for the networks
to performance-dependently evolve with continuous growth and contraction to achieve a structure of specific size for enhanced information processing.

These distinct network growth strategies were also compared with the Erd\"os-Renyi random networks (N0). 
The size of a random network yielding the same level of performance varies over a much larger range. The average number of nodes needed for successful performance for a random network is markedly more than that of an evolved network. However, though a majority of small random networks will not yield the prescribed performance, since the spread in sizes is large (Fig.\ref{fig3}), there exist certain small random networks that can perform efficiently. The search for these special random networks is hard, and there is no over-arching principle to guide the search in the extremely large space of network realizations. This is one of the drawbacks of reservoirs comprised of random networks. This observation leads to a prompt question, why do the majority of the ER random networks fail to efficiently solve the benchmark tasks even after having as many nodes as compared to the performance-dependently evolved networks? Furthermore, is there any similarity between the evolved networks with those few ER random networks that managed to, if not perfectly but reasonably well solve the tasks?

\subsection{Network growth vis-a-vis random chance: emergent performance-dependent network structure}

Graph theoretic properties of the networks such as density and node-degrees are evaluated to understand the similarities and differences between the networks generated by the diverse growth strategies (N1, N2 and N3) with that of Erd\"os-Renyi random networks (N0).

\begin{gather}
\begin{aligned}
\delta = & M/N(N-1)
\end{aligned}
\label{eq_NetDensity}
\end{gather}

\begin{figure}[h]%
\centering
\includegraphics[width=0.99\textwidth]{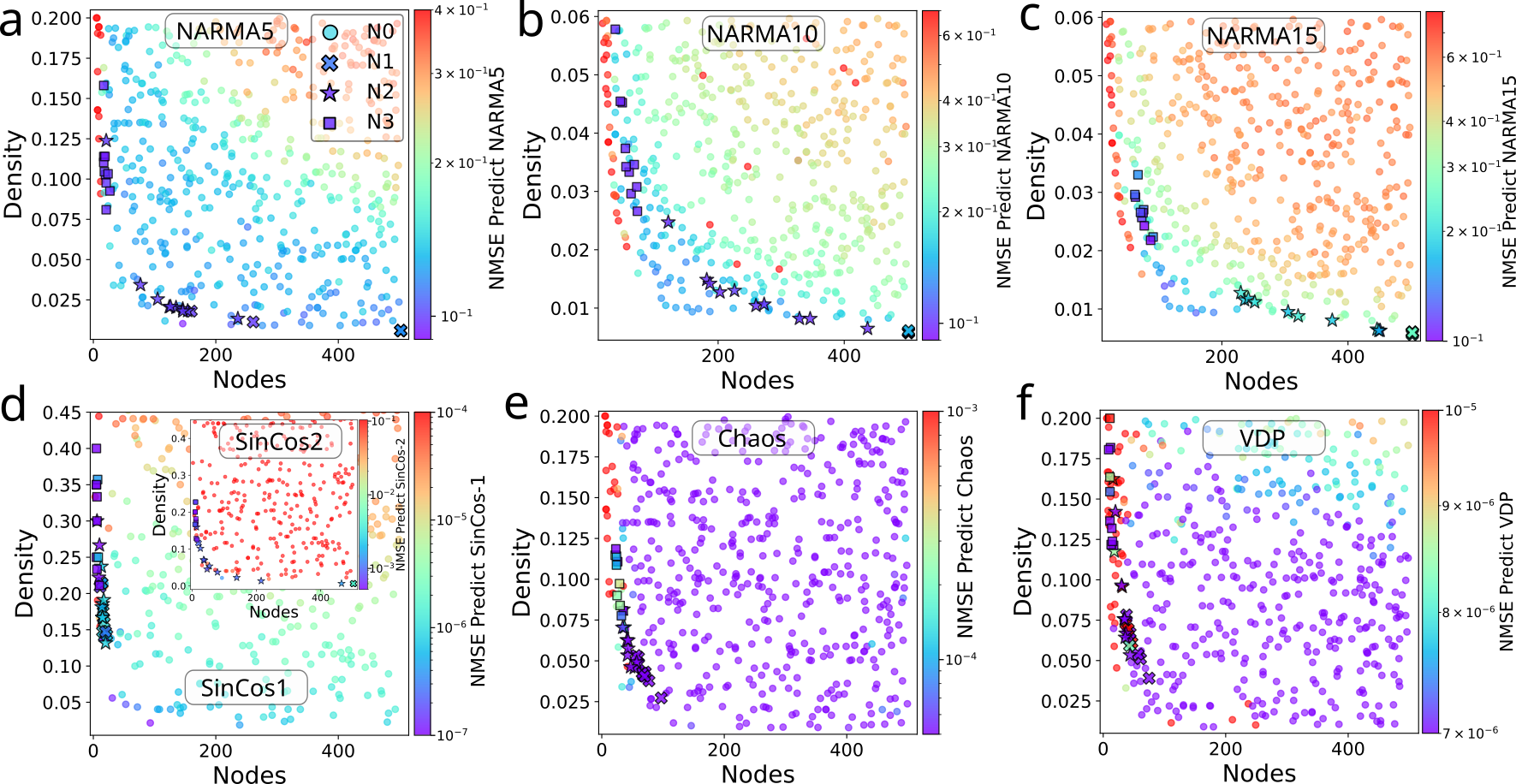}
\caption{\textbf{Organization of networks generated by distinct growth model in Nodes($N$)-Density($\delta$) space.} Densities of final generated networks are represented as a function of nodes (Eq.\ref{eq_NetDensity}) with symbols crosses, stars and squares respectively for models N1, N2 and N3. Circles covering the entire $N-\delta$ parameter space represent 500 Erd\"os-Renyi random networks (N0). The symbols are filled with colors corresponding to their performance for solving the tasks NARMA-5, 10 and 15, SinCos-1 (inset SinCos-2), Chaotic trajectory and periodic orbit from Van der Pol oscillator respectively in panels (a) to (f). }\label{fig4}
\end{figure}

The density ($\delta$) of a directed network is given by Eq.\ref{eq_NetDensity}, where $M$ is the total number of links in the network \cite{Newman2010}. Densities of networks generated by the three different network growth models for each of the benchmark tasks are plotted with respect to the nodes in that network in Fig.~\ref{fig4}. The color of each symbol represents the performance of the generated network for that task, where the purple color exhibits the best while red represents the worst performance for the corresponding task. Furthermore, 500 Erd\"os-Renyi random networks (N0) are also plotted along with their performances on the $N-\delta$ space in Fig.~\ref{fig4} for the respective tasks. The random networks are generated to have nodes and densities in the range corresponding to that of different tasks.

The final networks generated by the distinct growth strategies (N1, N2 and N3) are perfectly aligned along a rectangular hyperbolic curve, reflecting an inverse relationship between density and nodes. Interestingly, in the space of all possible random network reservoirs, the ones that yield the best performance, as exemplified by low errors, lie in the vicinity of the $\delta-N$ scaling function of the networks generated by distinct growth models. This feature is exhibited by the presence of the deep blue symbols for N0 (reflecting low errors) {\em only} on the curve displaying the inverse proportionality of density with respect to number of nodes.

\begin{figure}[h]%
\centering
\includegraphics[width=0.99\textwidth]{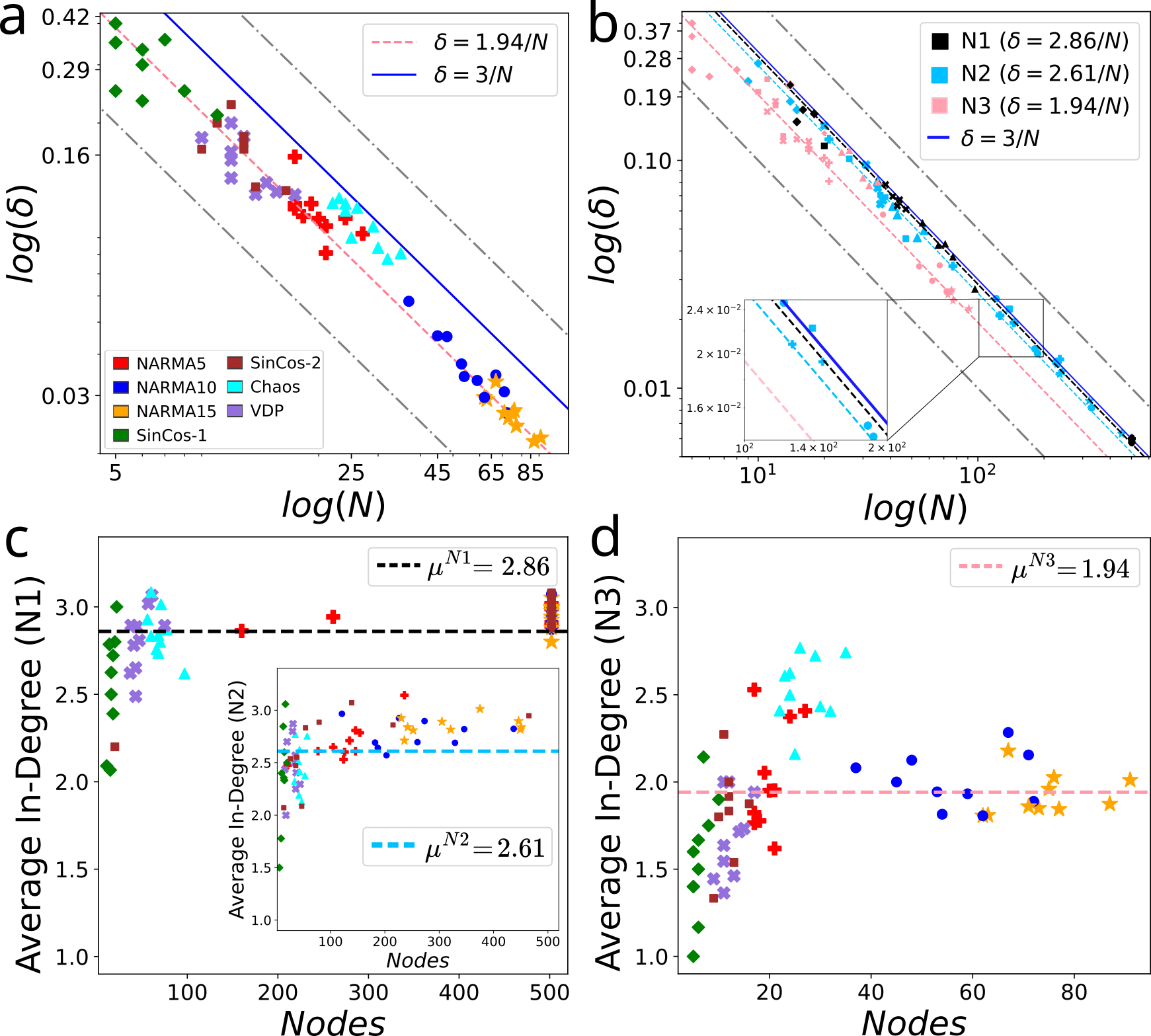}
\caption{\textbf{$N-\delta$ scaling over different network growth strategies.} (a) The densities ($\delta$) of the final network N3 evolved from performance-dependent node addition and deletion, represented against the number of nodes ($N$), over $log-log$ scale, for 7 different tasks. (b) Comparison of the scaling of the densities as a function of number of nodes, for different network growth models, N1 (black symbols), N2 (sky-blue symbols) and N3 (pink symbols). The lower and upper grey dotted-dashed lines in panels (a) and (b) represent the limiting cases of network growth with $\delta_{min}=1/N$ and $\delta_{max}=5/N$, respectively. Average in-degrees of the final networks with respect to the nodes for different tasks are shown for model N1 in (c), N2 in (c, inset) and for N3 in (d) for different tasks. The mean of all the average in-degrees ($\mu^{N1}, \mu^{N2}$ and $\mu^{N3}$) across different tasks for that particular model are represented with the dashed lines in panels (c) and (d).}\label{fig5}
\end{figure}

Another aspect of elucidating the observed $\delta-N$ scaling of the networks generated by the different growth-based models is in terms of the degrees of the nodes. The number of edges ($M$) in a growing network can also be expressed as a product of the average degree of the network ($\mu$) and nodes ($N$) i.e $M=\mu N$. Plugging it back to Eq.\ref{eq_NetDensity} reveals that the density of the growing or evolving networks scale as $\delta\approx\mu/N$. When a network grows, each new node can make $M^{New}\in[M^{New}_{min}, ..., M^{New}_{max}]$ number of new edges. This means that over several growth iterations, the network will have its average degree equal to $\overline{M^{New}}$ or $\mu=\overline{M^{New}}$. Therefore, the density of a growing network should scale as $\delta=\overline{M^{New}}/N$ shown with a navy-blue line in $log(\delta)-log(N)$ in Fig.\ref{fig5}(a) and (b). $M^{New}_{min}, M^{New}_{max}$ and $\overline{M^{New}}$ are 1, 5 and 3 respectively in this study. Furthermore, such a growth will always stay bounded by $\delta_{min}=M^{New}_{min}/N$ and $\delta_{max}=M^{New}_{max}/N$, both are represented by grey dotted-dashed lines in Fig.\ref{fig5}(a) and (b). In the case of N1 model, the average degree of the generated networks $\mu^{N1}\rightarrow\overline{M^{New}}$ (Fig.\ref{fig5}(c)), for large networks. This implies that there is no preference for the degree of the added nodes, and all degrees are chosen with equal probability, leading to a simple statistically averaged quantity.
 
The only small difference between $\mu^{N1}$ and $\overline{M^{New}}$ appears because of the lower average degrees of the very small networks obtained for the SinCos-1 task. Since this task is not complex enough, the networks converge rapidly within a few iterations (Fig.~\ref{fig2}(a) and Fig.~\ref{fig3}(d)) implying that new nodes are added with fewer new edges i.e $m<\overline{M^{New}}$.

However, the networks evolved by model N3 have average (in-)degree $\mu^{N3}<\overline{M^{New}}$ (Fig.\ref{fig5} (d)). This explains why all the evolved networks fall below the $\delta=\overline{M^{New}}/N$ (blue) line in Fig.\ref{fig5} (a). In other words, these networks are specifically evolved by the N3 model to have lower node degrees in order to achieve enhanced computational capacity. This shows another interesting feature of the networks generated by the performance-dependent network evolution strategy. 

 In summary, the significant features of the networks evolved with different strategies vis-a-vis uninformed growth and the random network is as follows: The first feature is that the average degree is smallest when networks evolved with the strategy of performance-dependent node addition and removal. That is, the networks that yield the best performance are sparser as compared to the ones with poor performance generated by alternate growth strategies. The second important feature is that the networks that emerge under performance-dependent node addition, as well as removal, have not only the lowest pre-factor in the scaling relation, but they also converge to low network size $N$. This is clearly evident in the fact that N3 networks occupy the low $N$ end of the hyperbolic arm of the scaling function in Fig.\ref{fig3}.

\subsection{Asymmetric distribution of input and readout nodes facilitate enhanced information processing}

In the traditional reservoir computing framework, it is a common practice to randomly select any number of input-receiving as well as output (or readout) nodes from the pool of $N$ reservoir nodes. However, there is no criterion for the selection and number of input and output nodes for efficient information transmission, processing, and reading-out. Therefore, all the distinct network growth models (N1, N2 and N3) described in the sections above are equipped with a degree of freedom for the selection of input ($I$) and output ($O$) nodes. A new node added by the \textit{node addition module A} can also be an input node with probability $P^{I}$, and it can independently also be an output node with probability $P^{O}$. We set $P^{I}=P^{O}=0.5$ in this study, meaning that a new node attaching to the network will independently have a 50\% chance of being an input or output node. There is a 25\% chance that this new node is both an input-receiving as well as a readout node, and from here onwards such nodes are denoted as \textit{Common Nodes} ($C=I\cap O$). Also, there is a 25\% chance that a new node is neither an input nor an output node, and these nodes are denoted as \textit{Unique Nodes} ($U=N - (I\cup O$)).

\begin{figure}[h]
\centering
\includegraphics[width=0.99\textwidth]{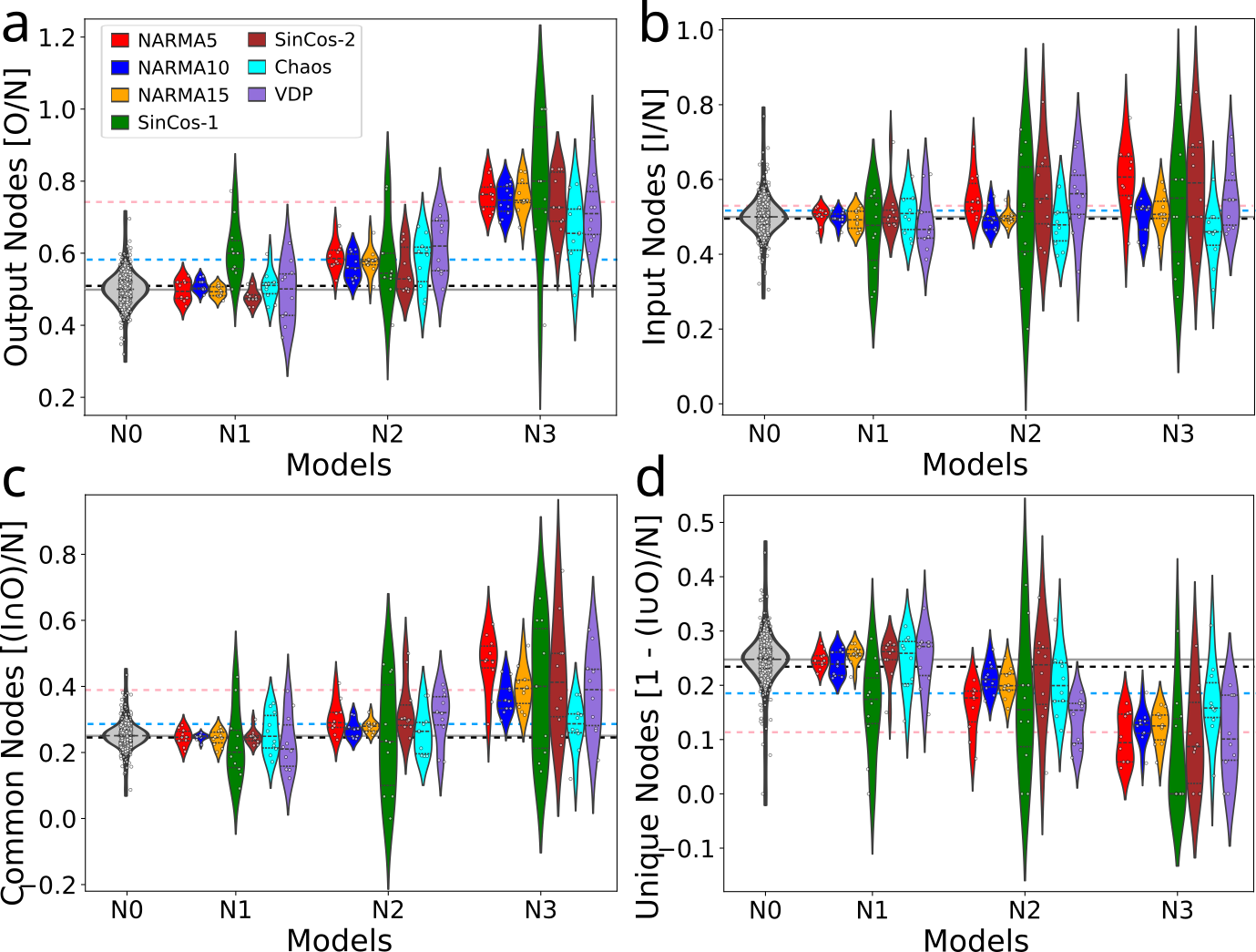}
\caption{\textbf{Asymmetric distribution of input receiving and output (readout) nodes in performance-dependently evolved networks.} Fraction of output ($O/N$) (a), input ($I/N$) (b), common (both input and output nodes, $C/N$) (c) and unique network nodes ($U/N$) (d) are compared for different network growth models (N1, N2, and N3) along with that of Erd\"os-Renyi random networks (N0) for reference. The mean over all the tasks for each of the fractions is represented with colored lines for the models N0 (solid grey), N1 (dashed black), N2 (dashed sky-blue) and N3 (dashed pink).}\label{fig6}
\end{figure}

The final networks generated by the performance-independent node addition (N1) model exhibit an equal fraction of input ($I/N$) and output nodes ($O/N$) for all the tasks, as shown in Fig.\ref{fig6} (a) and (b) respectively. Consequently, the fractions of \textit{Common} ($C/N$) and \textit{Unique} ($U/N$) nodes are normally distributed around 0.25, as shown in Fig.\ref{fig6} (c) and (d) respectively. These results were expected from the N1 model because the networks grow without the deletion of any nodes and also without any feedback for the selection of specific nodes. As a reference, the $I$ and $O$ nodes in the Erd\"os-Renyi random networks (N0) are also selected with the same values of the probabilities $P^{I}$ and $P^{O}$ respectively. The means of all these fractions over different tasks for the Erd\"os-Renyi random networks (N0) (grey solid line) are always close to that of N1 networks (black dashed line in Fig.\ref{fig6}). Therefore, the fraction of $I$, $O$, $C$ and $U$ nodes of the Erd\"os-Renyi networks also have the statistically same distributions as that of N1 networks. 

However, the fraction of readout nodes in the networks evolved with performance-dependency (N3) is surprisingly high as compared to that of N0 and N1 networks. The number of $O$ nodes is consistently high over all the different benchmark tasks. The mean fraction of output nodes $(O/N)$ over all the tasks is $\sim0.74$ (Fig.\ref{fig6} (a)). The mean fraction of input nodes $(I/N)$ is $\sim0.53$ which is just a slight shift as compared to a 50\% rise in output nodes. Interestingly, the fraction of \textit{Common} nodes ($C/N$) rose to $\sim40\%$. This means that not only the readout nodes are rising but the nodes that are both $I$ as well as $O$ are getting selected when networks are evolving with performance-dependent node addition and removal (N3). As a result, the fraction of \textit{Unique} nodes ($U/N$) falls to $\sim0.1$ on average for all the tasks. 

Furthermore, the networks generated by the N2 model also have a higher percentage of $O$ and $C$ nodes, that is $\sim58\%$ and $\sim29\%$ respectively. These fractions are in-between those of N3 and N1/N0 models. Since the N2 model represents partial evolution where networks can only grow depending on their performance without deletion, therefore, this slight rise in the fraction of $O$ and $C$ nodes was expected. 

This is an important finding that shows that the performance-dependent network evolution is shifting the readout nodes to a higher fraction that eventually enhances its computation capabilities. This shift toward the more readout nodes means that the ansatz vector scape for the linear regression becomes larger and richer increasing the overall accuracy. The network size prescribes how complex the latent space dimensions can be where the larger networks reflect the more complex dynamics of the reservoirs and the readout node number indicates how many of those states are essential for the output generation. Such an asymmetric distribution of readout and common nodes is therefore a unique feature that is required in networks to efficiently solve a given task with limited resources, i.e. number of nodes. This shift is a causal effect of performance-dependently addition and removal of nodes in an evolving network contrary to all different network growth strategies as well as Erd\"os-Renyi random networks.

\section*{Discussion}

We have proposed the concept of performance-dependent network evolution for efficient information processing and demonstrated that complex dynamical tasks can be efficiently and accurately solved with smaller evolved networks. Specifically, we showed, through the execution of a range of tasks of varying levels of complexity, that evolved networks exhibit enhanced computation capacity when compared to the Erd\"os-Renyi random networks. Furthermore, evolution through performance-dependent node addition and deletion outperformed all the networks generated with alternate network growth strategies, that are either uninformed or allow only performance-dependent node addition. This demonstrates the importance of performance dependency, together with continuous adaptive expansion and contraction of the network, to achieve an optimal size for its elevated performance. A prior study by P. François explored oscillatory responses by the biochemical networks with specific biochemical dynamics on evolving phenotypic networks \cite{PaulFrancois1, PaulFrancois2}. However, it was confined to smaller biochemical modules and lacked graph-theoretic measures for complex tasks. 

Performance-dependently evolved networks exhibit several unique features over their course of evolution such as a saturation in the growth of the network size, as compared to the linear and sub-linear growths observed in the other inefficient network growth strategies. The performance-based node addition/deletion evolution tends to generate sparser networks by preferentially expanding the networks with the nodes having lower degrees while deleting the ones otherwise. The densities of all the networks are found to follow a scaling in the $N-\delta$ space, irrespective of their performance dependency. Furthermore, performance-dependent evolution leads to a unique organization of the networks in the $N-\delta$ space, according to different task complexity. The evolved networks of simpler processes such as Sine-Cosine tasks are smaller and denser, and therefore are present on the top end of the scaling curve in contrast to the ones evolved for complex tasks such as NARMA-10 and 15, which have more nodes and occur on the lower end with low density. Significantly, this correlation offers a new approach to measuring the complexity of a computational task, based on the network size necessary for requisite performance and its corresponding density. 

The asymmetric distribution of input-receiving and readout nodes in the evolved reservoir network is yet another unique feature of the performance-dependent network evolution, with these networks having a high percentage of readout nodes and common nodes as compared to the input-receiving ones. This is an important finding suggesting the extraction of maximal and sufficient information from a network with a high fraction of readout nodes, which is contrary to the common practice of random readout node selection. Such an emergent asymmetric distribution of $I$ and $O$ nodes has been observed in biological networks as well, which naturally evolve for more complex processes. For instance, GRNs across different species have been observed to have a high fraction of readout nodes as compared to the \textit{recurrent} part of the networks \cite{RC_Jordi}.

The results presented here also shed light on the evolutionary path and scaling followed by the biological networks shown to have \textit{recurrent} structures to carry out biological computations. This general framework can be further extended to understand the formation of specific network structures for biologically relevant processes, node dynamics and evolutionary constraints such as mutations and spatial restrictions.

The unique features of the task-specifically and performance-dependently evolved networks hold promise for the machine-learning community to efficiently solve various classes of problems with smaller, specialized networks. Furthermore, recall that a few Erd\"os-Renyi random networks also managed to solve some tasks with satisfactory performance. However, since the majority of the random networks failed to solve the tasks, therefore generating efficient random networks is another painstaking task in itself. This search is especially onerous when one does not know either the network density or the fraction of readout nodes that need to be selected. Our results precisely provide pointers to those few Erd\"os-Renyi random networks that yield satisfactory performance by pinning their location in the vicinity of $\delta-N$ scaling of evolved networks. This enormously narrows down the search space offering a recipe for generating high-performing "designer" random networks. The asymmetric distribution of input-receiving and readout nodes can be further explored to enhance the efficiency of such Erdős-Rényi networks.

The performance-dependent network evolution framework presented here is a first-of-its-kind and comprehensive approach that addresses critical questions regarding efficient information processing networks. It offers insights from graph theory, complex dynamical systems, and machine learning perspectives, providing a formal description of process-specific network \textit{structure-function} dependencies and emergent scaling laws in evolving networks. This framework facilitates the generation of minimal and task-specific networks while elucidating their resulting graph-theoretic properties and scaling laws for efficient information processing. Our exploration lays the groundwork for advancing the field of network science by uncovering guiding principles for generating minimal and efficient task-specific networks and understanding their unique emergent properties.



\bibliography{sample}



\section*{Acknowledgements}

This work was supported by the Deutsche Forschungsgemeinschaft (DFG, German Research Foundation) under the Special Priority Program (SPP 2353, project number 501847579). S. Sinha acknowledges support from the J.C. Bose National Fellowship (Grant No. JBR/2020/000004).

\section*{Author contributions statement}

M.Y. conceptualized the study, performed simulations and prepared the figures. M.Y., S.S. and M.S. interpreted the results and contributed equally in writing the manuscript. 

\section*{Competing interests}
The authors declare no competing interests.





\newpage
\section*{Figures and legends}

\end{document}